\documentclass[12pt,oneside,a4paper,english]{article} % for Science, Engineering and Humanities and Social Sciences articles
\usepackage{url,hyperref,lineno,microtype,subcaption}
\usepackage[onehalfspacing]{setspace}
\usepackage{graphicx}
\usepackage{braket}
\usepackage{subcaption}
\usepackage{amsfonts,amsmath,amssymb,amsthm,nicefrac}
\usepackage{stringstrings}
\usepackage{mathtools} 
\usepackage{etoolbox}
\usepackage[margin=1in]{geometry}
\DeclarePairedDelimiterX{\abs}[1]{\lvert}{\rvert}{\ifblank{#1}{{}\cdot{}}{#1}}
\usepackage{fancyhdr}
\usepackage{lastpage}
\usepackage[numbers]{natbib}
\def\firstAuthorLast{} %use et al only if is more than 1 author
\def\Authors{James L. Webb\,$^{1,*}$, Luca Troise\,$^{1}$, Nikolaj W. Hansen\,$^{2}$, Jocelyn Achard\,$^{3}$, Ovidiu Brinza\,$^{3}$, Robert Staacke\,$^{4}$, Michael Kieschnick\,$^{4}$, Jan Meijer\,$^{4}$, Jean-Fran\c{c}ois Perrier\,$^{2}$, Kirstine Berg S{\o}rensen\,$^{1}$, Alexander Huck\,$^{1}$ and Ulrik Lund Andersen\,$^{1}$}
\begin{document}

\title{Optimisation of a diamond nitrogen vacancy centre magnetometer for sensing of biological signals} 
\maketitle

\author[\firstAuthorLast ]{\Authors} %This field will be automatically populated
\normalsize

\begin{abstract}

%%% Leave the Abstract empty if your article does not require one, please see the Summary Table for full details. MAX 350 words
\section{}
Sensing of signals from biological processes, such as action potential propagation in nerves, are essential for clinical diagnosis and basic understanding of physiology. Sensing can be performed electrically by placing sensor probes near or inside a living specimen or dissected tissue using well established electrophysiology techniques. However, these electrical probe techniques have poor spatial resolution and cannot easily access tissue deep within a living subject, in particular within the brain. An alternative approach is to detect the magnetic field induced by the passage of the electrical signal, giving the equivalent readout without direct electrical contact. Such measurements are performed today using bulky and expensive superconducting sensors with poor spatial resolution. An alternative is to use nitrogen vacancy (NV) centres in diamond that promise biocompatibilty and high sensitivity without cryogenic cooling. In this work we present advances in biomagnetometry using NV centres, demonstrating magnetic field sensitivity of approximately 100 pT/$\sqrt{Hz}$ in the DC/low frequency range using a setup designed for biological measurements. Biocompatibility of the setup with a living sample (mouse brain slice) is studied and optimized, and we show work toward sensitivity improvements using a pulsed magnetometry scheme. In addition to the bulk magnetometry study, systematic artifacts in NV-ensemble widefield fluorescence imaging are investigated. 
\end{abstract}

\section{Introduction}

Many biological processes generate electrical signals, for example synaptic transmission and muscular contraction. Such signals give key information on the functioning on biological systems, either for clinical diagnostic purposes (such as electrocardiography) or for fundamental understanding of processes and structure. Measuring these signals is typically performed using electrical probes carefully positioned in the desired region. This however poses difficulties if this region is not easily accessible, such as inside the brain, or where signals must be highly spatially resolved. An alternative approach is to measure the magnetic field produced when these signals propagate as electrical current. Existing techniques for this such as magnetoencephalography (MEG) or magnetocardiography (MCG) are limited by reliance on superconducting quantum interference device (SQUID) systems \citep{Krber2016}. These suffer a number of disadvantages including the need for an expensive magnetically shielded room and bulky cryogenic cooling, increasing the sample-sensor distance and resulting in weaker magnetic field at the sensor and poor spatial resolution.   

A number of technologies are currently being researched in an effort to replace SQUIDs and overcome these difficulties, including microelectromechanical (MEMs) sensors \citep{Li2017}, sensors based on magnetoresistive effects for magnetomyography \citep{Zuo2020} and in particular atomic (optically pumped) magnetometers for which cryogen-free MEG has recently been demonstrated \citep{Boto2018}. Another alternative in the solid state is to sense magnetic fields using nitrogen vacancy (NV) centres in diamond\citep{Taylor2008}. Such defects, specifically the NV$^{-}$ center, have an electronic structure that results in magnetic field dependent variation in fluorescence output under illumination with resonant microwaves (MW) and 532nm green laser light. Although as yet the sensitivity to magnetic field of NV sensors is worse than SQUIDs or atomic magnetometers, they have a number of advantages well suited to biosensing. They operate under ambient conditions, from DC to kHz+ bandwidth needed for ms-timescale biological signals and in close proximity (or even within \citep{McGuinness2011}) cells or tissue, due to the excellent biocompatibility of diamond. The high density of such defects in the material also makes possible imaging of biological signals near the optical diffraction limit with micrometer scale resolution, in a way not possible by competing techniques. Finally, the high dynamic range of NV sensors means they can be operated without extensive magnetic shielding from ambient magnetic noise (e.g. from 50/60Hz mains) and unlike atomic magnetometers, they do not require shielding from the Earth's magnetic field to achieve maximum sensitivity. 

Figure \ref{fig:fr6},a) shows the simplified level structure of an NV$^{-}$ defect in diamond used for such sensing. Electrons excited by a 532nm pump laser can decay in two ways. The most likely, labeled (1) is directly in the triplet state between ground $^3A_2$ and excited $^3$E state. However, if microwaves (MW) are applied on resonance between the ground m$_s$=0 $^3A_2$ state and its fine m$_s$=$\pm$1 (and hyperfine) split levels (2.8GHz) followed by pump excitation, decay back to the ground state can occur through process (2), via singlet shelving states $^1$A, $^1$E and nonradiative processes/weak infrared emission peaking at 1042nm\citep{Acosta2010}. This leads to a drop in fluorescence output on microwave resonance of up to 30\% for a single NV, or 1-2\% for a large NV ensemble\citep{Rondin2014}. This process is termed ODMR: optically detected magnetic resonance. The m$_{s}$=$\pm$1 states can be Zeeman split by a magnetic field, which shifts the resonance frequency, allowing detection of the field amplitude by monitoring fluorescence output. A 1nT field produces a shift of 28Hz when the field is aligned along an NV axis and shifts down to a few Hz can be detected within one second of acquisition time, corresponding to the picotesla level fields expected from biological signals. 

Biosensing can cover bulk magnetometry, collecting all emitted fluorescence from a diamond and aiming for maximum overall sensitivity. It also covers widefield imaging magnetometry, aiming for realtime imaging of magnetic field at high resolution using a microscope objective and a camera. It can also cover scanning magnetometry using single or few NVs, via confocal microscopy, a scanning NV tip or nanodiamonds within a biological sample\citep{Schirhagl2014}. In this work we focus only the first two areas, bulk and widefield sensing. An excellent up to date summary of the field in general is given in the recent work by Barry et al.\citep{bar32141}.

To date, bulk magnetometry biosensing has proved challenging. Barry et al.\citep{Barry14133} demonstrated detection of single neuron action potentials from a marine worm, with excellent 15pT/$\sqrt{Hz}$ low frequency sensitivity. Few groups have yet to replicate this level of sensitivity, let alone on a biological specimen. We note a distinction between DC-low frequency sensitivity and the typically significantly better AC ($>$10kHz) sensitivity achievable using pulsed dynamical decoupling techniques, where 10pT/$\sqrt{Hz}$ has been demonstrated \citep{Hall2010, Wolf2015,Masuyama2018}. These frequencies are however too high for detection of many biological signals in the sub-kHz range. The best reported low frequency/DC sensitivities are typically worse\citep{Zheng2019,zhe65436,Clevenson2015} with few-nT/$\sqrt{Hz}$. Vector magnetometry has also been demonstrated in the DC-low frequency range, important since magnetic fields from complex biosystems may not be easily directed along a single sensitive NV axis\citep{Zhao2019,Schloss2018}.

For widefield imaging, a goal is to image the field from biological signals (such as neuron action potentials)\citep{Hall2012,Hall2013}. However, the pixel imaging sensitivity is on the order of 250nT-2$\mu$T/$\sqrt{Hz}$, as yet insufficient for imaging pT-level fields\citep{Karadas2018}. Biosensing imaging work has therefore focused on samples which have a strong magnetic field due to the presence of ferromagnetic material\citep{LeSage2013, Glenn2015, Davis2018}.  Several works have studied magnetic microbeads, the tracking of which can have applications in biology\citep{Kazi2019}. The majority of work however has focused on imaging non-biological samples with far stronger magnetic fields, such as geological samples\citep{Glenn2017}, vorticies in superconductors \citep{Schlussel2018} and ferromagnetic nanowires\citep{Lee2018}. 

In this work we present experimental measurements using bulk magnetometry and widefield sensing. We demonstrate a magnetometer setup with 100pT/$\sqrt{Hz}$ sensitivity which is designed and optimised for measurements of magnetic field from biological samples. We show the setup, in an inverted microscope geometry, is also well suited for widefield imaging. We discuss the adaptations and considerations required for biological magnetometry and the potential limitations and pitfalls in terms of sample damage and imaging artifacts. Finally, we discuss the future prospects for these methods, with reference to our measurements using pulsed magnetometry. 
\\

\section{Materials and Methods}

\begin{figure}[ht]
\begin{center}
\includegraphics[width=\textwidth]{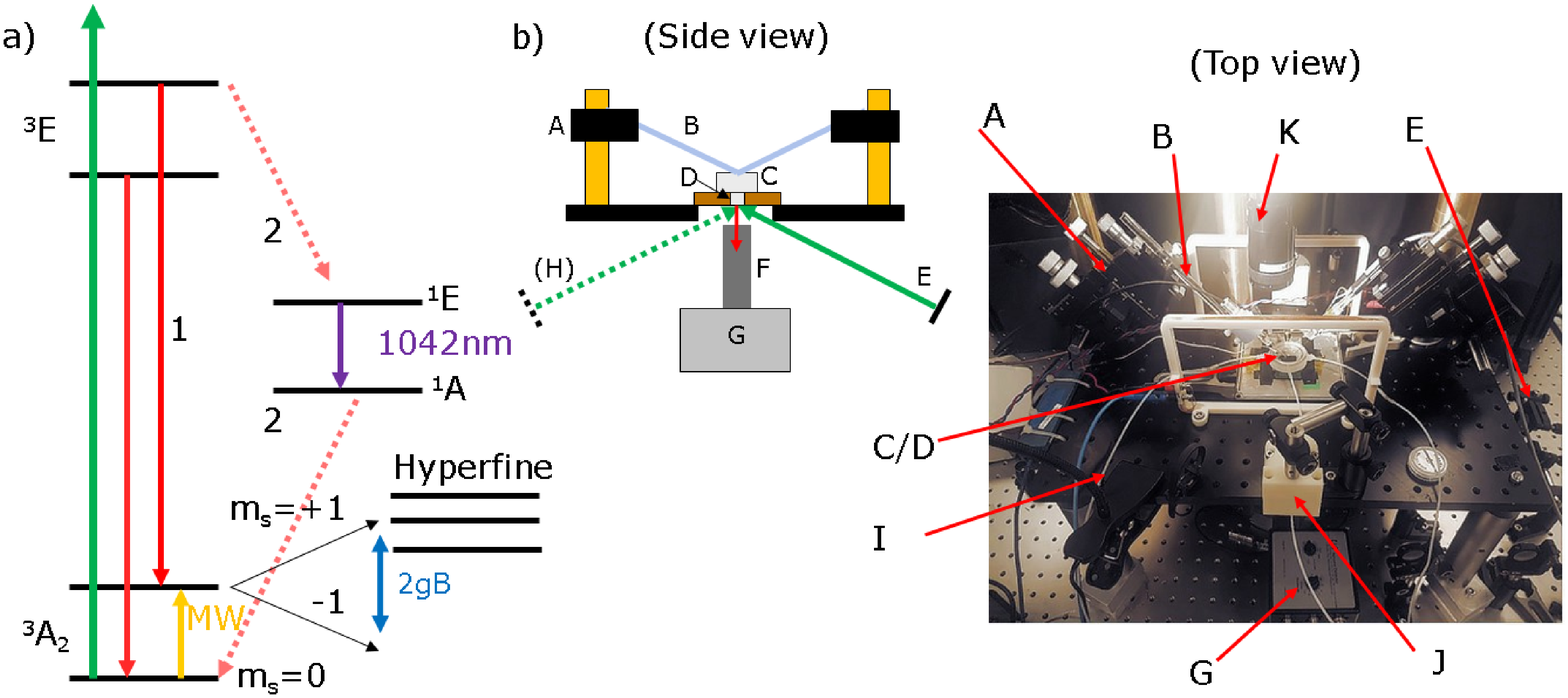}% This is a *.eps file
\end{center}
\caption{a) Simplified level structure of an NV$^{-}$ defect in diamond. Under 532nm green laser illumination and on microwave resonance between the fine (and hyperfine) split levels, decay back to the $^3$A$_2$ ground state can occur nonradiatively/via IR emission, resulting in magnetic field sensitive drop in red fluorescence. b) Simplified schematic and photograph of our electrophysiology microscopy-style setup, with illumination and fluorescence collection under a raised platform, with access from above by contacting electrodes and a white light microscope. Key: A) micromanipulators, B) stimulation/readout electrodes, C) sample chamber, D) diamond (in AlN plate, on MW antenna board), E) tilted mirror directing laser to sample, F) collection lens/microscope objective, G) photodetector, H) optional tilted mirror for second laser (not shown in image), I) solution feed tubing, J) DC offset permanent magnets, K) white light microscope.  }
\label{fig:fr6}
\end{figure}

Our NV magnetometer setup as pictured in a schematic and photograph in Figure \ref{fig:fr6},b). We use an inverted microscope geometry, where laser excitation and light collection is performed with the diamond held on a raised platform, above a microwave antenna board and below a custom-3D printed bath chamber for holding a biological sample (C and D, Figure \ref{fig:fr6}), with the solution fed via capillary tubes (I) from a peristaltic pump. This configuration has developed from typical electrophysiology microscopy setups. The setup allows a test sample to be placed directly above the diamond, with easy access to the sample from above, allowing the entry of stimulating electrodes using 2-axis adjustable micromanipulators (A,B) and a white light microscope (K) for positioning the electrodes and sample examination. 

Fluorescence from the diamond can be collected from beneath the platform (F), using either a microscope objective or a condenser lens in a beam tube. Via a mirror, this is then directed to the photodetector (G), placed below and infront of the raised platform. Using a beamsplitter, a reference beam is also supplied to the detector for noise rejection. Pump light is supplied from a tilted mirror (E) to the right of the setup at Brewster's angle for diamond (67.5 degrees) to maximise coupling into the diamond (light can also be supplied from the left using a second tilted mirror (H) from a second laser if required). Polarisation of the pump light is controlled using a polarising beamsplitter and half wave plate. Alternatively, both pump illumination and fluorescence collection can happen through the same lens using a dichroic mirror and optical filter to split the red/green components. This allows a focused beam to be supplied for imaging. In order to split the defect energy levels, optimise sensitivity in a single on-NV axis direction and obtain good contrast and sensitivity, a mT scale DC offset field is provided by two neodymium permanent magnets (J) placed infront and behind the sample chamber along with field coils for applying test signals. To switch between bulk sensing and imaging, minimal simple adaptation is required: the condenser lens/beamtube (F) is replaced with a microscope objective using the same screw fitting and the balanced photodetector (G) replaced with a suitable camera.  
\\

\subsection{Diamond Material}

\begin{figure}[ht]
\begin{center}
\includegraphics[width=8.5cm]{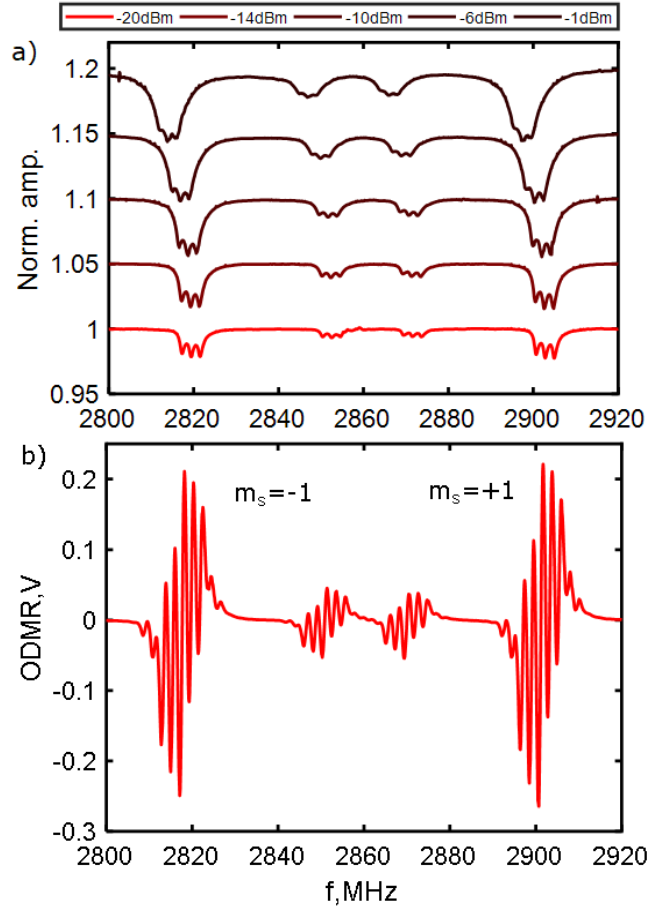}% This is a *.eps file
\end{center}
\caption{ODMR from our diamond sample. The $^{14}$N diamond was $^{12}$C enriched with 20$\mu$m of CVD overgrowth, then proton irradiated with N concentration $\approx$5ppm. a) Normalised change in DC photovoltage with 2W pump power (max 6.5mW fluorescence) at a range of microwave input powers, each offset by 0.05. A $\approx$1.6mT offset field was applied with fine adjustment from field coils to overlap 2 NV axes and boost contrast to a maximum of 5.1\%. Due to microwave broadening and resulting loss of the hyperfine features, actual peak sensitivity is reached at a lower 3.8\% contrast. The downward frequency shift observed is due to heating by the microwave field. b) ODMR at peak sensitivity using a 33kHz MW frequency modulation of 500kHz mixed with a 2.16MHz hyperfine transition drive (see method details in the work by El-Ella et al.\citep{ElElla2017}). This boosts sensitivity to 100pT/$\sqrt{Hz}$. Linewidth FWHM is approximately 1MHz.}
\label{fig:fr5}
\end{figure}

Careful selection of exposure dose, $^{12}$C isotopic purification to remove the negative effects of $^{13}$C spins \citep{Jahnke2012, 2004.01746} and understanding the role of strain\citep{Trusheim2016} are key to maximising sensitivity by achieving narrow microwave resonance linewidths and high fluorescence contrast (change in output between microwaves on/off).  In this work we use a diamond (grown at LSPM, Paris) with a 20$\mu$m CVD-grown $^{12}$C purified layer, doped with 5ppm nitrogen-14, carefully irradiated using H$^{+}$ ions (at U. Leipzig) and subsequently annealed at 800$^{\circ}$C. Figure \ref{fig:fr5} shows ODMR for our $^{12}$C purified diamond. The linewidth FWHM, proportional to the dephasing time T$_2$$^{*}$\citep{Bauch2018}, is on the order of 1MHz, as compared 10-50MHz for high density and unpurified samples.  

Peak sensitivity balances high NV number (high brightness) and ODMR contrast and linewidth. Brighter samples with higher NV density tend to have a lower contrast and broader linewidth and those with high contrast and narrow linewidths tend to be low in brightness. Further discussion of the sensitivity limitations, including the effects of laser line narrowing and microwave power broadening are given in the work by Dreau et al. \citep{Drau2011}. Our diamond has a linewidth comparable to the better diamonds in literature (FWHM$\leq$1MHz), contrast of 1-2\% at maximum field sensitivity and total fluorescence collection of 5-6mW at 2W of pump power. Based on these values, a simple estimate based on realistic collection efficiencies gives a shot noise limited sensitivity to DC and low frequency fields on the order of 10-20pT/$\sqrt{Hz}$. 
\\

\subsection{Laser coupling and light collection}

A factor critical to good performance in diamond NV magnetometry is coupling laser light in at Brewster's angle for diamond. This produces significant improvements in fluorescence generation by both coupling more pump light into the diamond and increasing internal reflection of the pump light, illuminating more NV centres\citep{Bougas2018}. This also has the additional benefit of minimising damaging pump laser leakage into any biological sample. This configuration, used for all experiments in this work, is achieved in our setup using a tilted mirror, directing the light in from the right hand side of the setup. Using an extra dichroic mirror, our setup also allows pump light to be directed through a microscope objective, with light incident perpendicular to the sample. Due to the limited field of view, this however can lead to saturation in the ODMR amplitude and reduction in contrast and sensitivity at higher power. For imaging we therefore use a microscope objective just for fluorescence collection. For the Brewster's angle configuration, we rotated the polarisation of the incident light using a half wave plate to ensure maximum transmission into the diamond. 

Our setup was also designed to utilise more than one laser beam incident on the diamond at the correct angle, using an additional tilted mirror to the left hand side of the setup.  Although laser light from two separate sources is not coherent, both beams can be sampled into the reference of a balanced photodetector in order to effectively reject the technical noise from both lasers simultaneously. This permits using 2 lower cost DPSS 1-2W lasers in combination to achieve greater sensitivity, as compared to having to purchase a single, high power, high stability laser. For imaging, our collection efficiency was limited by the numerical aperture of the objective. For bulk magnetometry, we used an aspheric condenser lens (Thorlabs, antireflective coated) to maximise fluorescence collection. 
\\

\subsection{Detection and Sensing Limitations}

Sensing with an NV centre ensemble is fundamentally limited by shot noise from the strong fluorescence background originating from transition directly from excited $^3$E to ground $^3A_2$ triplet state and the low 1-3\% change in fluorescence level on microwave resonance. This is the difficult task of detection of a small dip in brightness on a large bright background with a consequently high shot noise level. This shot noise limited regime means sensitivity improves with the square root of the number of NV centres (larger ensemble) and by high pump laser power (up to the order of Watts). In order to address the NV$^{-}$ defects and reach these power levels, we used a 532nm DPSS laser (Coherent Verdi) of maximum power output 2W. We note that the majority of the fluorescence produced is trapped within the diamond by its high refractive index. 

In practice, reaching this shot noise limited regime is difficult due to laser technical noise and electrical readout noise. In our setup, we rejected laser technical noise using a balanced photodetector (Newport Nirvana) or by sampling the input laser beam and digitally subtracting it from the fluorescence signal\citep{Schloss2018}. The majority of our electrical readout noise originates primarily from the transimpedance amplifiers within the detector and the choice of photodiode. For the balanced detector, we used the inbuilt photodiodes, with bandwidth limited by the electric balance feedback circuit to 100kHz. For the alternative sampling and subtraction method, we chose a photodetector with a larger area photodiode and higher capacitance (Thorlabs DET100) to ensure maximum fluorescence detection, at the cost of higher noise. Detection bandwidth was 10MHz, more than sufficient for kHz signal detection. 

For the majority of this work sensing was performed by supplying microwaves frequency modulated at 33kHz with 500kHz modulation width. This modulation could then be detected in the collected fluorescence from the diamond and the ODMR spectrum recovered by sweeping microwave frequency and performing detection using a lock-in amplifier. Once the ODMR spectrum had been obtained, a fixed microwave frequency was used corresponding to the point of maximum slope and thus giving the maximum response to any magnetic field induced change in fluorescence output. We used a three-frequency drive method as outlined in \citep{ElElla2017}. Measurements were performed using a continuous wave method with constant microwave and laser power. 

In the section \textit{Pulsed Measurement} of this work, short microwave and laser pulses were instead used for sensing, measuring the change in output on the readout laser pulse with and without a prior microwave pulse. Pulses were generated from a TTL pulse generator (Spincore Pulseblaster) and fed to fast RF switches (Minicircuits) and an acousto-optic modulator (Isomet). The objective of the pulsed scheme was to reduce microwave power broadening effects and enhance ODMR contrast\citep{Taylor2008,Rondin2014}.
\\

\subsection{Measurement Environment}

In order to maintain optimum sensitivity, it is necessary to well control the environment in proximity to the diamond. This is difficult to achieve in practice due to the need to hold biological samples in solution that can vary in volume or composition over time, due to movement (particularly muscle tissue) and due to changes in temperature, all of which can shift the optimum microwave frequency and power required to maximise magnetometer sensitivity.

The efficiency of microwave coupling to the diamond can be considerably affected by having a water or other solution in close proximity. This is particularly critical for antennas with a narrow resonance, such as wire loops\cite{Sasaki2016}. To counter this, we fabricated a custom-designed broadband antenna on a PCB board to be placed adjacent to the diamond. Using this antenna, we could reoptimise MW power (often by several orders of magnitude) after a stable carbogenated solution level sufficient to hold the specimen within was reached. Typically a depth of 0.5-1cm was used, defined by the need to fully cover and contain within the solution a biological sample, such as a tissue slice or a mouse muscle. In order to maintain sensitivity, it was critical to keep a constant solution level and constant flow (provided by a peristaltic pump), avoiding shifts in the MW resonance. Stabilisation of the level could be passively achieved by careful custom chamber design for the pumping rate required,  by manually constricting a section of tubing to reduce inflow or by controlling the rate of inflow or outflow using needle valves in the feed and return piping. These could either be manually operated or controlled from a level sensor on the chamber. If necessary to fully inhibit sample movement, inhibitor chemicals as butanedione monoxime (BDM) (dissected tissue) or local anasthetic (living specimen) could be introduced to the solution feed and pumped to the sample.  

Another factor was gradual thermal drift, caused by microwave or (as previously discussed) laser heating. Thermal drift could, over the course of minutes, push the optimum microwave resonance frequency away from the setpoint frequency, reducing sensitivity. To correct for this, we designed automated software to perform an ODMR sweep every few measurements, with the computer automatically determining the point of maximum field sensitivity and maintaining the microwave frequency at this point. This could alternatively be performed continuously by keeping the frequency at a point that maximised the strength of a low frequency (111Hz) signal applied to test coils aligned with the field sensitive axis. 
\\

\subsection{Sample-Diamond Separation}

Since magnetic field strength drops as the inverse of distance (current carrying wire approximation) or the cube of the distance (magnetic dipole approximation), only a small difference in separation between diamond and biological sample (order of $\mu$m\citep{Karadas2018}) can make a significant difference in signal detection strength. This poses particular challenges for biological systems, where the tissue generating the signal may unavoidably be many millimeters from the sample (such as within muscle or under bone) or cannot be directly in thermal contact with a diamond heated by a laser beam. In our setup, we used a reflector/insulator layer above the diamond (aluminium foil/Kapton tape) to keep the sample as close to the diamond as possible while minimising heat transport from diamond to sample. We also found that samples could float from the surface away from the diamond.  Using a Pt and nylon harp (standard equipment for electrophysiology) helped stop this without disrupting the DC offset magnetic field. We also used Ti hooks and electrodes to hold the sample onto the diamond, with care taken not to damage the sample. 
\\

\subsection{Stimulating Electrode Type and Design}

In order to stimulate a biological response, our setup was configured to use standard electrophysiology probes attached to micromanipulators, which could supply current pulses to the sample. In electrophysiology experiments, only the electrical performance of the test electrodes is important and no offset magnetic field is used. This means many types of electrical stimulation electrodes that can be purchased contain magnetisable steel, often in the outer sheath of concentric-type electrodes. Such electrodes become magnetised in the DC offset field, which disrupts the field close to the diamond, significantly reducing magnetometer performance. To avoid this problem we developed an alternative, which was to use custom-made electrodes using Pt/Ir wire with a glass or plastic outer sheath. We found similar electrodes could be made from biocompatible, non-ferrous Ti or W wire (Cu was not biocompatible unless coated with Au or Ag/AgCl). 
\\

\subsection{Averaging}

Due to the weakness of biological signals, it is necessary to repeat the signal N times and then average, with improvement in sensitivity proportional to 1/$\sqrt{N}$. This however places considerable demands on a biological sample, for example due to fatigue in a muscle, due to damage induced in the tissue by repeated stimulation or due to thermal degradation. We found that a bath of carbogenated (5\% carbon dioxide, 95\% oxygen) solution allowed the sample to survive for many hours, permitting many thousands of stimulations and averages. 
\\

\section{Results}

\subsection{Magnetic Noise}

In any practical laboratory or clinical setting, there will be substantial magnetic field generation from mains transformers in equipment and in building wiring. This occurs primarily at the fundamental mains frequency (European 50Hz or North American 60Hz) and the 3rd harmonic (150 or 180Hz) arising from magnetic hysteresis in the transformer, but also at higher harmonics and at multiples of 3 phase frequencies. Such low frequency fields are difficult to shield against, requiring full  mu-metal enclosure, impractical for processing of biological samples, or extremely costly fully shielded rooms into which feeding piping or cabling is equally hard. Active cancellation, by using field coils to generate a counter field to the noise, can reduce it but this is hard to achieve for multi-axis sensing at a small 1-2mm diamond, at kHz bandwidth and operating at sub-nT/$\sqrt{Hz}$ sensitivity. The majority of commercially available systems cannot achieve these levels of field cancellation. 

The alternative we present here is for the magnetically sensitive signal from the diamond magnetometer to be processed and filtered by Fast Fourier Transform (FFT) methods. This requires not only effective filtering but minimal filtering, since many biological signals have frequency components in the 10-500Hz range and any strong filtering here will distort or remove the desired signal entirely. Such a filter must adapt to any drift in mains frequency and be able to capture any transient noise (such as from a fan or pump turning on and off). Further, it must counter phase drift in the mains supply, which acts to spread the mains noise across a wider frequency range. 

\begin{figure}[ht]
\begin{center}
\includegraphics[width=\textwidth]{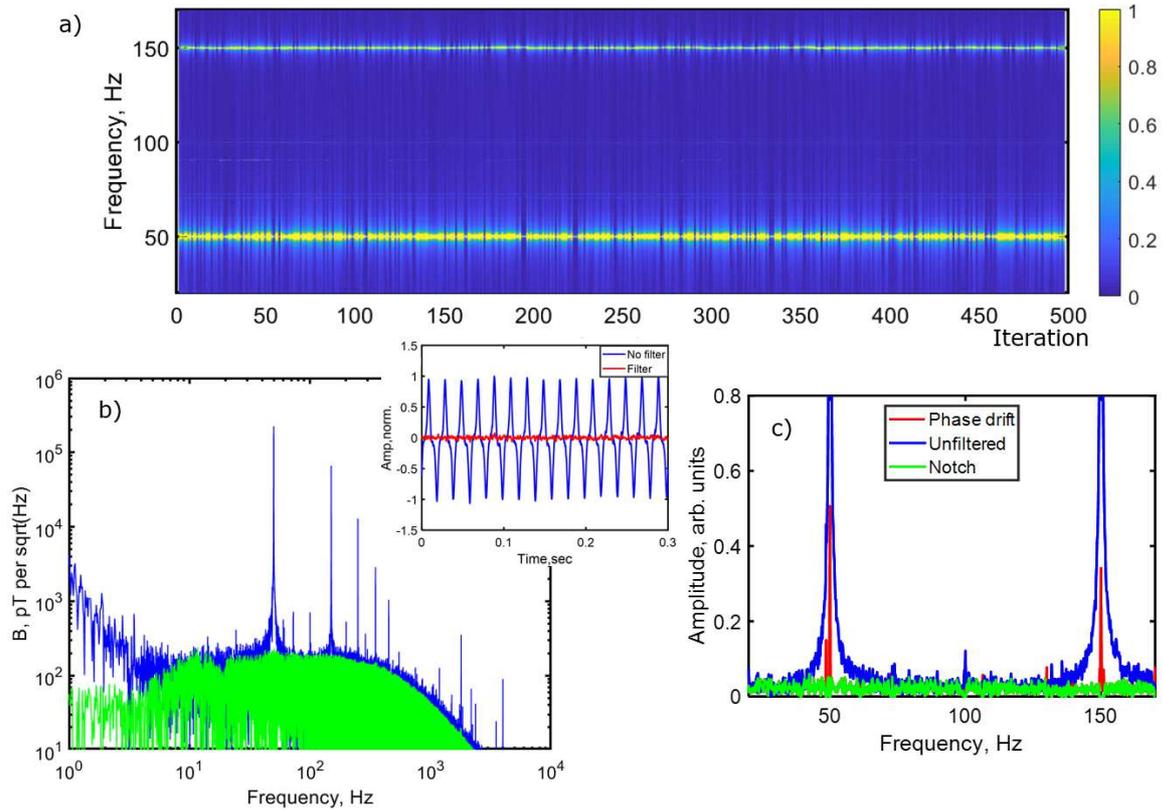}% This is a *.eps file
\end{center}
\caption{a) Normalised spectrogram of magnetometer pickup from 30-170Hz from FFT of successive 60sec long measurements (iterations) at regular intervals over a period of 8 hours. The variable phase drift during each measurement is seen as a broadening in frequency of the 50/150Hz mains noise peaks. b) Magnetometer sensitivity from a single 60sec measurement using 1W pump power while magnetically sensitive before (blue) and after (green) filtering. The low frequency noise floor is approximately 150-200pT/$\sqrt{Hz}$, (inset) timeseries of signal before and after filtering, showing significant improvement obtained, c) Single 60sec measurement before filtering (blue), after timeseries filtering (red) to remove phase drift and after notch filtering (green).}
\label{fig:noise}
\end{figure}

In Figure \ref{fig:noise} we demonstrate how effective this can be. Magnetometer data was collected for 60sec in order to give good frequency discrimination. Low frequency and DC drift arising from laser power fluctuations is first removed by a highpass filter at 10Hz. Next, a 50Hz zero phase pure sine wave is correlated with a readout of the fundamental mains frequency obtained from tapping input to a transformer, in order to obtain a time series of the phase drift over the measurement period. A slight constant phase shift can be added to this signal to account for any phase shift due to inductive effects, then normalised and subtracted from the magnetic readout. This process could be repeated several times to remove multiple (additive) sources of noise with different phase. It can then be repeated at 150Hz, 250Hz or other strong harmonics of mains. Consequently, the phase drift broadening is removed in the frequency domain. Narrow notch or bandstop filters can then be used, centered on the noise frequencies, to remove any remaining noise components (Figure \ref{fig:noise},c) . Any high frequency RF noise could then be removed by a final lowpass filter at 5-10kHz. The overall effect in the data is a reduction in noise by orders of magnitude, down to the electrical and optical noise floors (inset, Figure \ref{fig:noise},b).

The ability to remove strong background magnetic noise components is key to allowing diamond NV sensing to be used in an ordinary lab or clinical environment without extensive and costly magnetic shielding for applications in research and medical diagnosis (e.g. magnetocardiography). This work complements our recent work on a flexible handheld magnetometer, containing the diamond, focusing optics and electronic components\citep{Webb2019}. 
\\

\subsection{Laser Heating}

In order to optimise sensitivity in NV centre experiments limited by fluorescence shot noise, high laser powers of the order of several Watts are required. This presents a challenge of focusing a high power visible laser beam onto a mm-size diamond while remaining within the $\leq$37C range from most biological processes. We found that sudden changes in temperature, such as from fast ramping of the laser to high power, would damage biological samples. Heat must be dissipated from the diamond while minimising the distance between diamond and sample, to maximise magnetic field strength at the diamond. Samples often cannot be placed directly on a metal heatsink, as this would short out any conduction paths and potentially interfere with the microwave or DC magnetic offset fields. 

\begin{figure}[ht]
\begin{center}
\includegraphics[width=8.5cm]{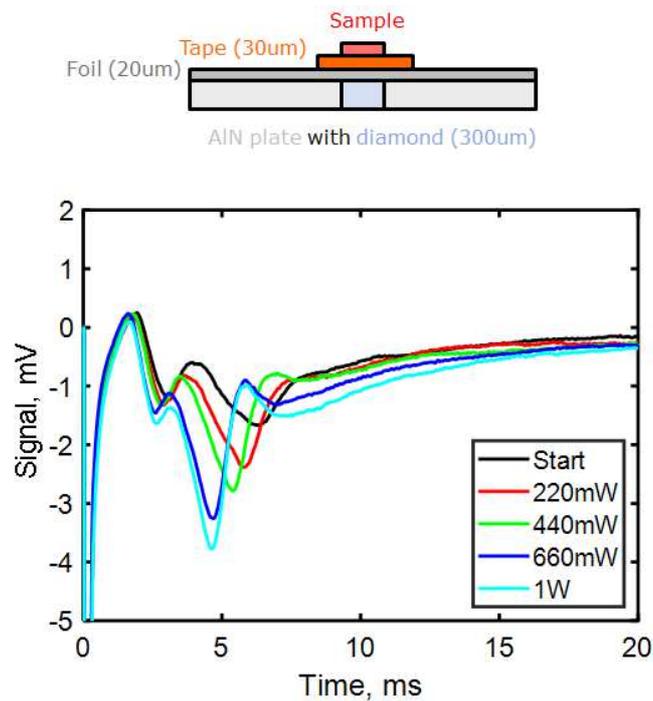}% This is a *.eps file
\end{center}
\caption{Change in mouse brain slice hippocampal field potential in the CA1 region at specific laser powers during laser ramp up (at 0.01W every 20s). The signal was recorded using a Pt/Ir electrical probe inserted into the region. Laser power was ramped to a maximum of 1W, with the slice thermally isolated from the diamond by a layer of 20$\mu$m thickness aluminium foil heatsunk into the surrounding solution and 30$\mu$m thickness Kapton tape. The diamond is within the center of a 2x2 cm AlN plate.}
\label{fig:fr444}
\end{figure}

Figure \ref{fig:fr444} shows the effect of slow ramping the laser power on the electrical readout on the local field potential evoked in the CA1 region of the hippocampus in a slice preparation of a mouse brain. Placing the sample directly onto the diamond surface, heated by the pump laser, was found to cause rapid damage or death to the slice. By using a metal foil layer as heatsink and to reflect stray laser light back into the diamond and thinner Kapton tape as electrical insulation, it was possible to ramp the laser slowly to high power while keeping the sample alive, with the signal strength slightly improving through gradual heating to a more optimal temperature for biological processes (35-37C). We observed no signal for direct slice transfer to a diamond preheated by the laser at 1 or 2W, attributable to thermal shock on placing it on the warm diamond. Ideally, the foil/Kapton layers could be replaced with thin glass or polymer (to be more robust against pressure from above) coated with Al to form a mirror. It should be noted that standard commercial glass coverslips are too thick for this purpose (type \#0 are 85-105$\mu$m thick).

One additional way of achieving good thermal dissipation is to have the diamond in contact with a SiC wafer\citep{Barry14133}. Another method we utilise in this work was to use plates of aluminium nitride (AlN) with a precision laser cut hole for the diamond. AlN is cheaper to purchase, easier to machine and has a good thermal conductivity while being an electrical insulator. This allowed heat to be dissipated out from the side facets of the diamond. We note from our previous work that thermal effects on the magnetometry can be canceled by driving two m$_s$=$\pm$1 hyperfine transitions, although at a factor 2x cost in sensitivity\citep{Wojciechowski2018}. 

Although we present electrophysiology measurements from the hippocampal slice here, we were unable to observe these signals in the magnetometer readout. Our previous estimates (detailed methodology in \citep{Karadas2018}) had indicated that reaching nT/$\sqrt{Hz}$ sensitivity along with signal filtering and averaging would be sufficient to resolve signals from a brain slice. We speculate that this failure was due to magnetic field cancellation at the diamond. We consider that this occurs due to current propagation in other directions in the slice than the desired direction we impose using the permanent magnet DC bias field to maximise the magnetic field along the most sensitive NV axis. We found it extremely difficult to position the slice (in a solution bath) correctly to direct the current in the hippocampus. Higher spatial resolution and/or vector magnetometry is required in order to probe the cause of the absence of magnetic signal. In our setup this could not be achieved without reducing ensemble size, fluorescence collection and excessively reducing sensitivity. Although  widefield imaging microscopy cannot resolve pT-level signals at the present time, this or scanning either the sample or the diamond on a probe over the biological sample may prove a method to overcome this issue in the future. 
\\

\subsection{Pulsed Measurement}

Substantial effort in recent years has gone into developing pulsed laser/microwave schemes for field detection and imaging. These have been very successful in improving sensitivity to high frequency AC fields ($>$10kHz) via dynamical decoupling methods, rather than in the limit of low frequency and DC signals. A significant problem for DC sensing is the time required to read a large ensemble of NV centres required to achieve a high level of sensitivity. This can limit the measurement bandwidth to below that required for many biological sensing applications, particularly if laser power must be limited in order to protect the biological sample from damage. 

\begin{figure}[ht]
\begin{center}
\includegraphics[width=8.5cm]{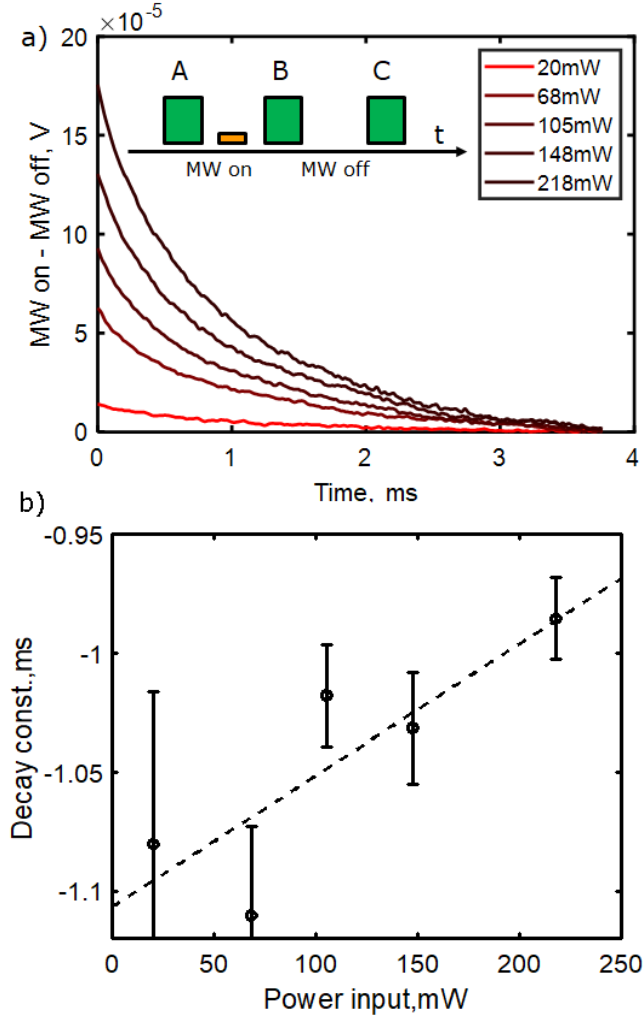}% This is a *.eps file
\end{center}
\caption{a) Difference in fluorescence vs readout pulse time (B,C). The system is initialised with a laser pulse A. Before B a 680ns $\pi$ microwave pulse is applied (MW on) and before C, no MW pulse is applied (MW off). The difference in phtotodetector output voltage during these two pulses as a function of time was measured and the difference is plotted in the figure (MW on-MW off, V). Data is taken at a range of pump laser powers from a Gaussian beam of waist 200$\mu$m incident at Brewster's angle. b) Fitting to an exponential with error estimates based on readout noise. Typical decay length was 1ms, reducing with higher pump power. This long readout time restricts field sensing bandwidth to 250Hz or 83Hz for a 3 pulse, common mode noise rejection scheme. Linear fit (dashed) $y=mx+c$ coefficients are $m$=5.5x10$^{-4}$ms/mW and $c$=-1.1ms.}
\label{fig:fr111}
\end{figure}

Here we measured the ensemble readout time for our diamond. We used a long (100ms) initialisation laser pulse (A) to ensure that as many NVs as possible were initialised in the m$_s$=0  ground state. We then sent a microwave pulse (on resonance at 2.91GHz) of variable length $\tau$ (MW on) This was followed by a second laser readout pulse (B), followed by a repeat of this sequence without the microwave pulse (MW off), finishing with a second readout pulse (C) (see schematic, Figure \ref{fig:fr111},a).  We measured the difference in fluorescence output between the readout pulse with MW on (A) and MW off (B) as a function of time, as would be necessary in a measurement scheme such as Ramsey interferometry. We varied the length of the microwave pulse $\tau$  to maximise this difference, which was found at $\tau$=680ns. For a single NV, this would correspond to a $\pi$-pulse. For our large ensemble, this corresponds to an approximate $\pi$-pulse for the most number (far from all) NVs.  Figure \ref{fig:fr111} shows the result of these measurements, measuring the difference in fluorescence (detected photovoltage) as a function of readout laser pulse time and averaging over 1000 repetitions of the pulse sequence. We find a readout (reinitialisation) time with a decay time on the order of 1ms, fitting using a single exponential function. We estimate the error bars on the exponential fit in Figure \ref{fig:fr111},b) based on the maximum deviation in fit parameter produced by the peak to peak readout noise on the photodetector voltage, which decreased with higher laser power (more fluorescence signal). Within error bounds, the decay time could be decreased by 200-300$\mu$s by increasing laser power from 20mW to 218mW, the maximum possible using this measurement configuration due to power limitations imposed by our acousto-optic modulator. We estimate the laser power intensity at the diamond to be 1.5kW/cm$^2$. Assuming a linear extrapolation to the higher power density used, our readout time is comparable to the 150$\mu$s in the work by Wolf et al.\citep{Wolf2015}. This is likely due to the reason given in their work, related to Gaussian spread of pump laser power, but also due to variations in strain and local field across the diamond and NV ensemble. 

We note that for a large ensemble, a constant amplitude and phase microwave pulse will not correctly address all NVs, either as a $\pi$-pulse in this example or as a $\pi/2$-pulse for a sensing scheme based on Ramsey interferometry\citep{bar32141}. A prospect to overcome this and enhance sensitivity in the pulsed regime may be microwave pulse shaping, using an arbitrary waveform generator and optimal control methods to boost readout fidelity\citep{Nbauer2015}. 
\\

\subsection{Widefield Imaging}

The camera limitations of NV centre widefield fluorescence imaging are discussed in detail in our recent work (Wojciechowski et al.\citep{Wojciechowski2018}). A key problem with NV ensemble widefield imaging is bit depth. The majority of digital cameras in general and used for microscopy work in only 8 or 10 bit depth (since the human eye cannot distinguish a greater color or monochrome palette). With a large background fluorescence brightness, this can result in the on MW resonance contrast being spread across too few digitisation levels. We therefore use more specialised 12 and 16 bit cameras (Imaging Development Systems GmbH) in order to avoid artifacts associated with low bit depth. An ideal camera would be one with an initial analog black level correction stage, that can spread the contrast across the full digitisation range.  

Such a high level of background brightness means shot noise will significantly exceed readout noise or dark noise on each pixel. Based on ODMR measurements and from technical specifications, we can estimate the best sensitivity possible. Assuming a detection area of 5x5mm$^2$ with 3.5$\mu$m$^2$ pixel size (approximately 2 Mpixels) and a diamond with a ODMR linewidth of 1MHz, contrast of 2\% and collecting a fluorescence output of $\approx$6mW, we calculate the best per pixel shot noise limited sensitivity of an imaging system to be approximately 50nT/$\sqrt{Hz}$. This presently restricts imaging of biological signals to specific, low bandwidth cases rather than imaging of pT-level signals such as those from neural activity. 

\begin{figure}[ht]
\begin{center}
\includegraphics[width=\textwidth]{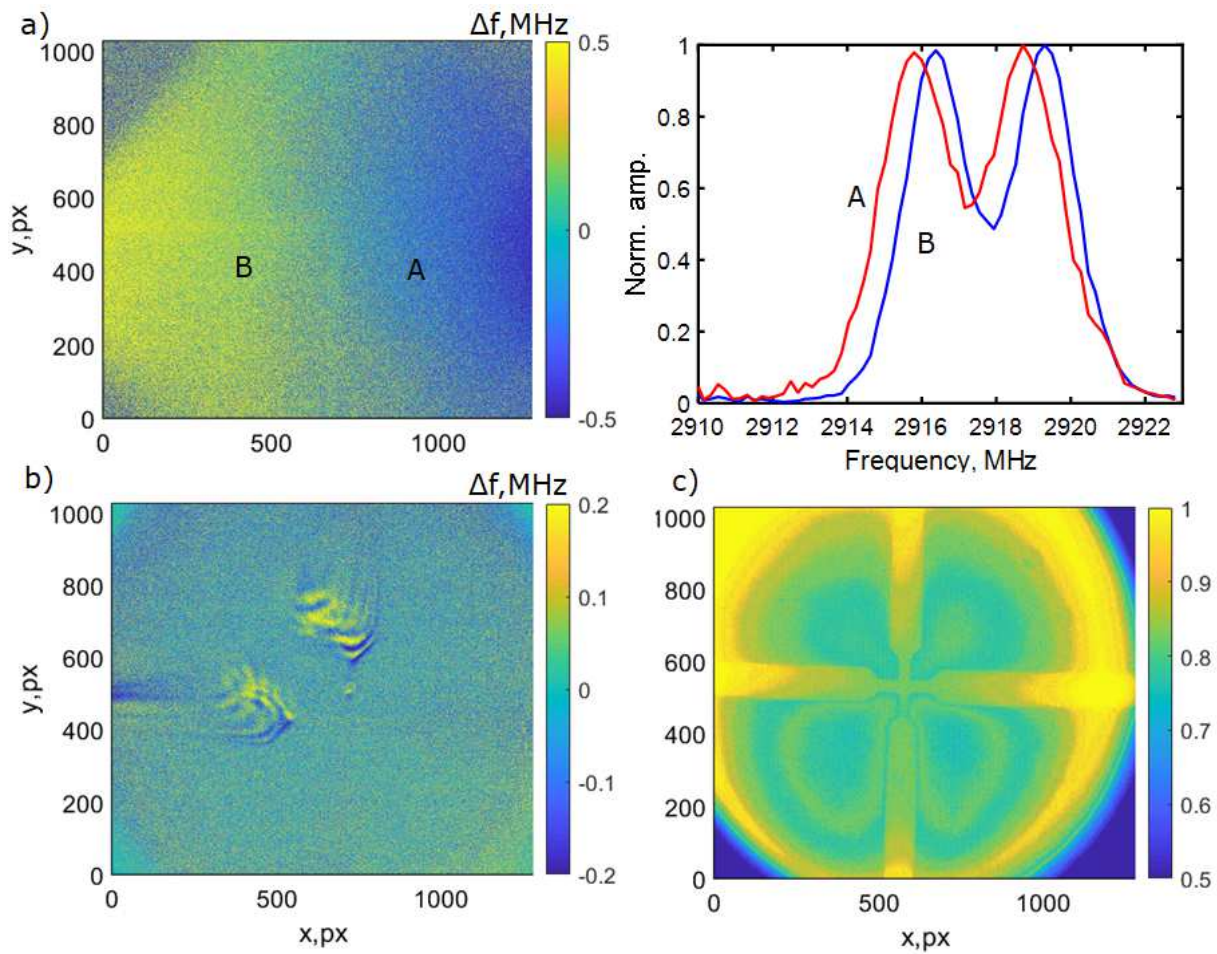}% This is a *.eps file
\end{center}
\caption{Artifacts in fluorescence imaging experiments of a Ti/Au 50nm cross on glass. In a) the variation of microwave field is shown across an image with a field of view (FOV) of approximately 1mm (using a x10 objective). The result is a difference in brightness at a given MW frequency that can resemble a field shift, as can be seen in the normalised ODMR plot averaging a 200x200pixel region at different places (A,B) on the image. b) The effect of vibrations (image wobble) on averaging of 100 successive images taken at 1ms intervals using a 100x objective with $\approx$100$\mu$m FOV. c) Imaging at 7Hz of vibrational modes of the diamond itself, which appear as slight changes in brightness that can resemble magnetic field patterns.}
\label{fig:fr778}
\end{figure}

A key issue we address in this work is imaging artifacts that can easily resemble magnetic fields. From our measurements, we demonstrate a number of these in Figure \ref{fig:fr778}. Variation in microwave power and resonance frequency across the image width can be more significant than the shift resulting from local changes in magnetic field (Figure \ref{fig:fr778},a), which can in turn be mistaken for localised magnetic field imaging. Jitter or wobble in repeated averaged images can produce patterns that appear to be localised to electrodes or areas where a field is expected (Figure \ref{fig:fr778},b). In addition, low frequency vibration of the diamond itself in the typical frequency range of biological signals can be observed in a magnetic-field like pattern of fluorescence brightness change (Fig \ref{fig:fr778},c). Finally, magnetic field-resembling diffraction patterns can arise from imperfect interfaces and air or solution gaps between diamond and sample.  
\\

\section{Discussion}

Improved sensitivity is key for biosensing, since this would allow greater diamond-sample separation while still being able to detect the signal, would allow more comprehensive measurement and filtering of background noise and would permit biological magnetic field imaging of even weaker signals (i.e. sub-pT to fT level for MEG). Substantial work in the field has focused on using pulse sequences to improve sensitivity\citep{bar32141}. This method appears challenging for DC/low frequency sensing, due to the bandwidth limitations of slow readout times for large ensembles, although microwave pulse shaping and optimal control techniques common in nuclear magnetic or electron spin resonance experiments (NMR/ESR) may help here.  A number of novel sensing techniques are currently being pursued which are suitable for sensing in this low frequency regime. This includes experiments using an optical cavity and infrared or green absorption\citep{Chatzidrosos2017,Ahmadi2018} or the very recent work by Fescenko et al.\citep{fes2897} promising sub-pT/$\sqrt{Hz}$ sensitivity using flux concentrators. Novel techniques at an early stage such as laser threshold magnetometry \citep{Jeske2016,Dumeige2019}  also appear promising. Such laser threshold experiments can potentially work with other materials beyond diamond, avoiding the problem of the large shot noise background from spin triplet fluorescence emission. Another possibility we suggest here is that pump light absorption may be exploited using interferometry techniques, such as by placing the diamond with the sample in one arm of a Mach-Zehnder configuration and detecting magnetic field through the effect of the field-dependent absorption of the diamond on the interference pattern generated at the output. This requires minimising reflective losses and good stable transmission through the diamond in order to be realised. 

It should be noted that many of these schemes pose problems for positioning the biological sample near the diamond yet out of the beam path. However, with sufficient sensitivity, the stand-off distance between the sample and the diamond can be increased (as is already necessary and possible for SQUID and atomic magnetometers), allowing it to be displaced laterally from the diamond while maintaining an orientation that directs the magnetic field along a sensitive NV axis. Many of these novel schemes are at an early stage and require further experiments to test their viability before applications to biosensing can be considered. 

In this work, we have demonstrated methods that avoid damage to biological samples. Further work is however required in terms of improving dissipation of heat away from the diamond and away from the sample. This is challenging since any heatsink must be thin to minimise sample-diamond separation and also electrically insulating. We note that there has been important parallel effort in constructing integrated minaturised sensors\citep{Ibrahim2019,Kim2019,Strner2019}. Although their reported sensitivity is several orders of magnitude worse, it is possible that by using techniques from semiconductor fabrication (particularly efficient heatsinking) that these devices may have a significant role to play in future NV biosensing.

Finally, it is hoped that advances in diamond material processing may drive future development. In particular, improvements in CVD growth, nitrogen levels and conversion to NV$^{-}$ centres, optimisation of irradiation and dealing with material strain. Patterning of the diamond may allow a greater fraction of the fluorescence to escape, boosting efficiency\citep{Huang2019}.
\\
	
\section{Conclusion}

Sensing of biological signals via their magnetic fields using diamond NV centers provides a potential route to better measure and understand them, whether in an electrophysiology-style microscopy configuration or from MEG-style sensing of inaccessible tissues in a living subject. Previous work has shown the technique to be biocompatible and widely applicable to a number of different systems, in both a bulk sensing and imaging configuration. There are, however, challenges in realising such measurements, both in general from material limitations and the need to filter out unwanted noise, to specific demands unique to biosensing, such as maintaining temperature stability. In the coming years, further advances in sensitivity are required in order to realise ambitious sensing goals. A number of promising ideas are currently under development, particularly using pulsed schemes and in improving diamond material quality.  

\section*{Conflict of Interest Statement}
%All financial, commercial or other relationships that might be perceived by the academic community as representing a potential conflict of interest must be disclosed. If no such relationship exists, authors will be asked to confirm the following statement: 
The authors declare that the research was conducted in the absence of any commercial or financial relationships that could be construed as a potential conflict of interest.

\section*{Author Contributions}
JLW, NWH and LT performed experiments for which data is shown as figures in this work. JLW wrote the manuscript with input from AH, KBS, ULA,JFMP and NWH. 

\section*{Funding}
The work presented here was funded by the Novo Nordisk foundation through the synergy grant bioQ and the bigQ Center funded by the Danish National Research Foundation (DNRF).

\section*{Acknowledgments}
We acknowledge the assistance of Kristian Hagsted Rasmussen for fabrication and diamond processing and Axel Thielscher, Mursel Karadas (current and former DTU Heath Tech) and Adam Wojciechowski (former DTU Physics, currently Jagiellonian University, Krakow) for contributions and prior experimental and theoretical modeling work. 

\bibliographystyle{apsrev} % for Science, Engineering and Humanities and Social Sciences articles, for Humanities and Social Sciences articles please include page numbers in the in-text citations
\bibliography{testarx}

\clearpage
\section{Figures}

\end{document}